\documentstyle[12pt,a41,epsf,psfig]{article}

\begin{document}

\begin{center}
{\LARGE\bf Recent developments of {\sc Pepsi} and {\sc Sphinx}}

\vspace{1cm}
{O.~Martin,  M.~Maul, A.~Sch\"afer}

\vspace*{1cm}
{\it Institut f\"ur Theoretische Physik, Universit\"at Regensburg,
D-93040 Regensburg, Germany}\\

\vspace*{2cm}

\end{center}

\begin{abstract}
Recent developments of the two MC-generators {\sc Sphinx} and {\sc
Pepsi} are discussed. In {\sc Sphinx} the process of polarized 
photo production was included to simulate the potential of a polarized
HERA to measure the photon structure function. In {\sc Pepsi} the
CC polarized cross sections on tree level have been added together with
the electroweak corrections to simulate their effect on
CC signals proposed for a polarized HERA.
\end{abstract}

\section{Introduction}

\vspace{1mm}
\noindent
The {\sc Pepsi} and {\sc Sphinx} programs are Monte Carlo event generators
for longitudinally polarized 
deep inelastic lepton-nucleon scattering (DIS) and longitudinally polarized 
high energy hadron-hadron collisions, respectively. 
They both use probabilistic methods to generate events with
unit weight.  The generation process can be split
up into three parts. At first, the kinetic variables of the partonic reaction
are determined according to the complete cross section yielding a two parton
initial state and a two, three or four parton final state. The 'partons'
include quarks, leptons and gauge bosons here. The next stage consists of
an iterative generation of initial and final state showers. 
Finally, the parton cascades fragment
into hadrons which may further decay. Until now, polarization is only
taken into account by the partonic cross section and the initial state shower.
A polarized treatment of the final state shower is in principle 
straight forward for longitudinal polarization, 
whereas to our knowledge no model of polarized fragmentation exists so far.

In the following we report recent developments concerning {\sc Sphinx} and
{\sc Pepsi}. For further information please refer to \cite{sphinx} and
the WWW pages
\newline
{\tt http://th.physik.uni-frankfurt.de/$\sim$martin/sphinx.html} and \newline
{\tt http://th.physik.uni-frankfurt.de/$\sim$maul/pepsi.html}.

\section{SPHINX}
\vspace{1mm}
\noindent
High energy scattering of a (quasi-)real photon on a hadron is very similar
to hadron-hadron collisions. 
The hadronic component of the photon is described by unpolarized and polarized 
photonic parton distribution functions much in the same fashion as for hadrons.
To leading order in the strong coupling constant one can thus clearly 
distinguish between a direct cross section and a resolved
cross section. In the first case, the photon couples completely to 
a parton of the hadron, in the second case it fluctuates into quarks 
(and possibly further into gluons) right before the hard interaction 
with the hadron occurs. To next-to-leading order this distinction breaks down,
but a new pragmatic definition can be found \cite{dirres}.

The unpolarized parton distributions of the photon have already been explored
at HERA during the last years and a resolved component of the total cross
section was established. Polarized photo production experiments have not
been conducted yet and therefore it is ``unknown territory''. We sought
to develop a tool for experimentalists which enables them to study
the feasibility of such experiments. 

{\sc Pythia~5.6} \cite{pythia}, the basis of the {\sc Sphinx} program, already
includes unpolarized photo production which made an extension of  {\sc Sphinx}
to longitudinally polarized photo production quite simple. In principle, only
the leading order polarized direct cross 
sections $f_i\gamma \rightarrow f_i\gamma$,
$f_i\gamma \rightarrow f_i g$ and $g \gamma \rightarrow q \bar q$ had to
be added and the  polarized parton distribution functions of the
photon had to be included. In the following we want to discuss some points of
interest:

\begin{figure}[hbt]
\begin{center}
\epsfysize10cm
\leavevmode\epsffile{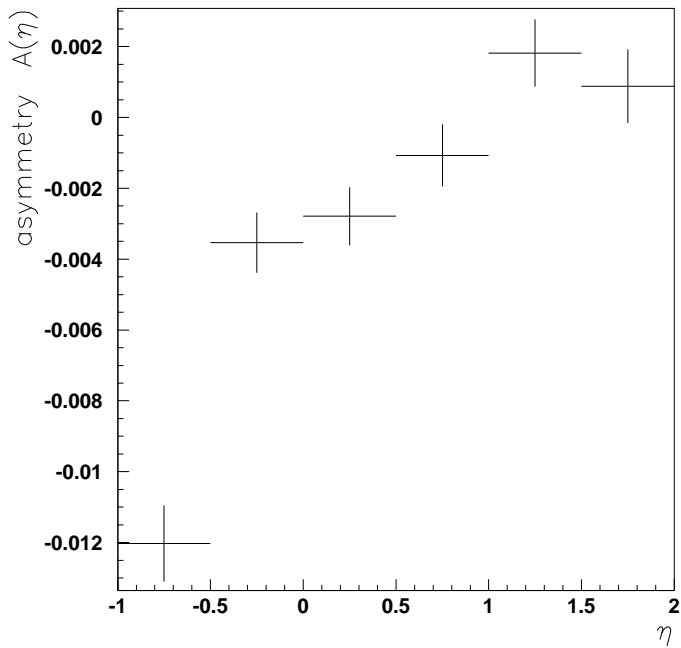}
\caption{\label{jets} Inclusive jet asymmetry for photo production
at the HERA polarized collider. {\em The error bars are not the expected
experimental (statistical) error bars but the error bars of
the Monte Carlo due to the limited number of simulated events.}}
\end{center}
\end{figure}

\begin{itemize}
\item The polarized parton distributions of the photon are completely unknown.
We included the minimally and maximally saturated 
ans\"atze by Gl\"uck and Vogelsang \cite{photonpdf}. 
These can be used to determine the
sensitivity of the asymmetries to the polarized parton distributions of the 
photon.

\item The current experimental setup at DESY is such, that electrons and
hadrons collide and react via the exchange of a virtual photon. The 
program therefore allows not only to generate photon-hadron events
at fixed energies but also electron-hadron collisions at fixed energies.
This is done by calculating parton distributions of the electron via
a convolution of the polarized Weizs\"acker-Williams spectrum with the
parton distributions of the photon.

\item The leading order cross sections for the production of QCD jets
with high transverse momentum and the production of direct photons
are implemented.

\item Double spin asymmetries for photo production are expected to be of the
order of one percent and below. Consequently, one needs at least
 $10^6$ {\it events
per bin} to achieve decent accuracy and program runs become somewhat lengthy.
This is illustrated by Figure~\ref{jets} which 
shows the inclusive jet asymmetry for photo production  at polarized HERA 
using the
maximally saturated set of polarized photon distribution function from 
\cite{photonpdf}. We
chose the same kinematic cuts as the authors of \cite{photoproduction}
 and the showering
routines were switched on.
$16\cdot 10^6$ events were simulated yielding a total execution time of
approximately 3 days on a Pentium~200 personal computer.
\end{itemize}

\section{PEPSI}
The possibility of measurements of CC events by missing momentum 
in deep inelastic
scattering is one of the most interesting perspectives of a polarized
HERA. For this purpose on twist-2 level within the
framework of the naive parton model the polarized contribution
to CC events together with the ${\cal O}(\alpha_{\rm em})$ electroweak corrections
is now included into the MC-generator {\sc Pepsi}. Unpolarized
CC events are already contained in the {\sc Lepto} code \cite{IER97},
which forms the basis of {\sc Pepsi},
and the
polarized extension can be provided on twist-2 level by a simple
substitution of the parton distributions.
For $W^-$ exchange:
\begin{eqnarray}
U(x,Q^2) & \longrightarrow & U(x,Q^2) + \lambda \Delta U(x,Q^2)
\nonumber \\
\bar U(x,Q^2) &\longrightarrow &  0 
\nonumber \\
D(x,Q^2) & \longrightarrow &  0  
\nonumber \\
\bar D(x,Q^2) & \longrightarrow & \bar D(x,Q^2) - \lambda \Delta \bar D(x,Q^2) 
\quad.
\end{eqnarray}
And for $W^+$ exchange:
\begin{eqnarray}
U(x,Q^2) & \longrightarrow & 0 
\nonumber \\
\bar U(x,Q^2) & \longrightarrow & \bar U(x,Q^2) - \lambda \Delta \bar U(x,Q^2)
\nonumber \\
 D(x,Q^2) & \longrightarrow & D(x,Q^2) + \lambda \Delta D(x,Q^2) 
\nonumber \\
\bar D(x,Q^2) &\longrightarrow & 0  
\quad.
\end{eqnarray}
\begin{figure}[hbt]
\begin{center}
\centerline{\psfig{figure=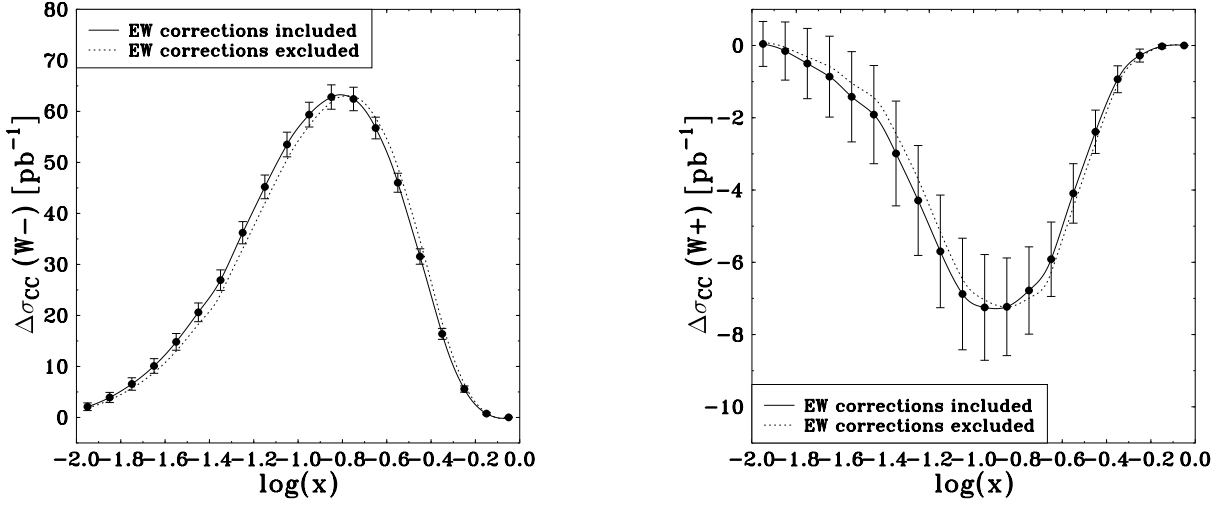,width=18cm}}
\caption{Spin dependent cross section $\Delta \sigma_{CC}:= 
\sigma_{CC}^{\lambda=1} - \sigma_{CC}^{\lambda=-1}$ in 
the case of $W^-$ exchange (left) and $W^+$ exchange (right). Solid line:
including electroweak corrections and dotted line: without them. The
error bars correspond to 25 pb${^{-1}}$ per relative polarization and
100 \% beam polarization. This corresponds to 200 pb${^{-1}}$ altogether
and 70 \% beam polarization for the electron and proton beam.}
\label{corr}
\end{center}
\end{figure}
$U$ denotes the u-like flavor of each generation, $D$ the corresponding
d-like flavor of each generation and $\bar U, \bar D$ are the corresponding
anti particles. $\lambda$ is the sign of the relative polarization configuration
of the two beams, i.e., $\lambda=1$ for anti parallel and $\lambda=-1$ for
parallel spin configuration. 
$x$ is the usual Bjorken $x$ and $Q^2$ the virtuality of
the exchanged W boson. 
The very important feature of the CC events is that the polarization
of the quarks can be expressed by their helicity. Consequently, for
the electroweak corrections on the twist-2 level we can use the corresponding
unpolarized but helicity dependent formula \cite{Bardin}, and include
those electroweak corrections again by a substitution of the parton
distributions:
\begin{equation}
q(x,Q^2) \longrightarrow \sum_{b=0,l,i,q} 
c_b\rho^2_C(P)\left[ \frac{1+P}{2} R_b + \frac{1-P}{2} \bar  R_b \right] \quad,
\end{equation}
with $P=P_l P_q, P_l,P_q = \{+1;-1\}$ for particles and anti particles,
and $c_b=\{1,Q_l^2,Q_lQ_q,Q_q^2\}, b=\{0,l,i,q\}$ being the charge factor.
$q$ denotes the quark flavor and $l$ the incoming lepton. 
The genuine electroweak
corrections are included in the form factor $\rho_C$ while the functions 
$R_b$ contain the one photon bremsstrahlung originating from the 
quark leg ($q$),
from the lepton leg ($l$) and from the interference of both ($i$). (0) denotes
the tree contribution. The functions $R_b$ are in general convolutions of the 
parton distributions, to be more precise
\begin{eqnarray}
R_0 &=& q(x,Q^2)
\nonumber \\
\bar R_0 &=& (1-y)^2 q(x,Q^2)
\nonumber \\
R_b &=& \frac{\alpha_{\rm em}}{\pi}\left\{ S_b q(x,Q^2) + \int_1^{1/x}
dz \left[ T_b \frac{r_b q(zx,Q^2) - q(x,Q^2)}{z-1} + U_b \frac{1}{z} q(zx,Q^2)
\right] \right\}
\nonumber \\
\bar R_b &=& 
\frac{\alpha_{\rm em}}{\pi}\left\{ \bar S_b q(x,Q^2) + \int_1^{1/x}
dz \left[ \bar T_b \frac{\bar r_b q(zx,Q^2) - q(x,Q^2)}{z-1} + 
\bar U_b \frac{1}{z} q(zx,Q^2)
\right] \right\} \quad.
\end{eqnarray}
\vspace{1mm}
\noindent
The exact definition of the functions $U_b,T_b,S_b,r_b$ and
$\bar U_b,\bar T_b,\bar S_b,\bar r_b$ is to be found in \cite{Bardin1}.
These formulae are available in FORTRAN in the program
{\sc Hector} \cite{Bluemlein}, and we took them partially from this
code. From the four contributions
the single photon bremsstrahlung radiation from the lepton leg
($b=l$) is dominant, while the others give only minor corrections.
\newline
\newline
In Fig.~\ref{corr} the electroweak (EW) corrections are shown
for the spin dependent charged current cross sections for
$W^-$ and $W^+$ exchange. The parton distributions used are from the 
polarized/unpolarized package GRSV LO STD described in 
\cite{GRSV}. 
In the kinematic range used here, i.e., 
$0.01<x,y<0.89$, $600 < Q^2$. The electroweak corrections are rather
small. Only in the case of 1000 pb ${^{-1}}$ there might be a chance
in the $W^-$ case that the error bars could become smaller than the
effect due to the electroweak corrections.

\section{Conclusions}
The new upgrade of the program {\sc Sphinx} including  
polarized photo production
allows the simulation of various models for the photon structure.
The upgrade on {\sc Pepsi} allows the analyses of polarized CC processes
at a polarized HERA including electroweak corrections. The effect of
the latter on the spin dependent part of the cross section is rather
small in the kinematic range covered by the polarized HERA.

\end{document}